\documentclass[aps,pra,10pt,a4paper,nosuperscriptaddress,twocolumn,showpacs,titlepage,showkeys]{revtex4}

\usepackage{times}
\usepackage{amsmath}
\usepackage{amssymb}
\usepackage{graphicx}
\usepackage[dvips]{hyperref}
\usepackage{subfigure}
\usepackage{booktabs}
\usepackage{rotating}

\begin{document}

\title{Entanglement of formation for qubit-qudit system\\
using partition of qudit sytem into a set of qubit system}
\author{Wen-Chao Qiang$^{a}$}
\thanks{Corresponding author.\\
E-mail address: qwcqj@163.com (Wen-Chao Qiang).}
\author{W. B. Cardoso$^b$}
\affiliation{$^a$Faculty of Science, Xi'an University of
Architecture and Technology, Xi'an, 710055, China\\
$^b$Instituto de F\'{\i}sica, Universidade Federal de Goi\'{a}s, 74.001-970, Goi\^{a}nia - GO, Brazil}

\pacs{03.67. -a, 03.65.Ud} \keywords{entanglement of formation,
qubit-qudit system,partition}

\begin{abstract}
Gerjuoy [Phys. Rev. A 67, 052308 (2003)] has derived a closed-  form
lower bound for the entanglement of formation of a mixed
qubit-qudit system (qudit system has $d$ levels with $d\geq3$).
In this paper, inspired by Gerjuoy's method, we propose a scheme that partitions a qubit-qudit system into
$d(d-1)/2$ qubit-qubit systems, which can be treated by all known
methods pertinent to qubit-qubit system. The method  is
demonstrated by a qubit-qudit system (The levels of qudit are
$d=3$ and $d=5$, respectively).
\end{abstract}

\maketitle

\section{INTRODUCTION}\label{s1}
Entanglement of quantum systems is an important physical resource
to realize quantum information tasks and quantum computation
~\cite{Nielsen}. The quantitative measure of entanglement is one
of the main research areas in quantum information theory and
quantum computation \cite{RH}, which have attracted much attention
of many researchers. In this sense, many useful measures were developed, such as: concurrence
~\cite{C1,C2,C3,C4,C5,C6,C7,C8,C9}, entanglement of
formation~(EOF) ~\cite{EF1,EF2,EF3,EF4,EF5,EF6}, geometric
measure~\cite{gm1,gm2,gm3,gm4}, entanglement witness
~\cite{W-1,W-2}, quantum discord ~\cite{discord_1,discord_2},
three-tangle~\cite {three_entangle}, etc. These
entanglement measurements are usually defined first for pure
states and then are extended to mixed states via the convex roof
construction. Because it requires complicated optimization
procedure, generally speaking, computing an entanglement
measurement for a given mixed quantum state is difficult. To the
best of our knowledge, only a few analytic expressions of
 some entanglement measurements for some special quantum
 systems, for example, the EOF  and geometric
measure of qubit-qubit states and isotropic states, are
obtained~\cite{C1,gm1,EF3,GV}. Some numerical algorithms for
computing some entanglement measures were also
developed~\cite{gm3}.

The concurrence and the EOF among all entanglement measurements
play an essential role due to some of other entanglement
measurements can be expressed by concurrence and the method
employed to derive analytic expressions of the EOF can be used to
derive analytic expressions of other entanglement measurements.
Wootters have obtained an elegant formula for qubit-qubit system
~\cite{C1}. Wei and Goldbart have also derived an analytic expression of the
geometric measure for two-qubit mixed states ~\cite{gm1}. Due to the fact of
the concurrence, as defined originally, is only suitable for qubit-qubit
systems and the optimization process to get analytic expressions for the EOF of a
general entangled state in higher-dimensional space to be complicated, several schemes were
proposed to find the lower bounds of the concurrence and the EOF
of general entanglement mixed states~\cite{gm1,C5,C6,C7}.
Using the Schmidt decomposition theorem~\cite{Nielsen},
Gerjuoy derived the lower bounds of the concurrence and the EOF of
any qubit-qudit system, and easily obtained Wootters' formula.
The method employs a set of $(2\times d)\times (2\times d)$ matrixes
$\textbf{S}^{ij}$, that are constructed via $\sigma_y^{(2)}\otimes\sigma_y^{(d)}$,
where $\sigma_y^{(2)}$ ($\sigma_y^{(d)}$) is the usual $\sigma_y$ Pauli matrixes
for the qubit (qudit) case. In this sense, in the present paper we propose a scheme
that partitions a qubit-qudit system into $d(d-1)/2$ qubit-qubit systems, which
can be treated by all known methods pertinent to qubit-qubit system.

This paper is organized as follows. In the next section, we
analyze Gerjuoy's method and propose our scheme, which
simplify Gerjuoy's procedure. In the section \ref{s2}, two
examples, qubit-qutrit ($d=3$) and qubit-qudit ( with $d=5$) systems, are given
to illustrate our scheme. The summary and discussion are given in
section \ref{s4}.

\section{ANALYSES OF Gefiuoy's SCHEME AND PARTITION OF
QUBIT-QUDIT SYSTEM INTO QUBIT-QUBIT SYSTEM}\label{s2}

To find the
lower bound of entanglement of formation (EOF) of the qubit-qudit
system, E. Gerjuoy \cite{C4} first defined $d(d-1)/2$ symmetric
square matrices $\textbf{S}^{ij},0\leq i \leq d-2$ and $j>i$,
whose elements all are zero, except for
\begin{subequations}
\begin{eqnarray}
  \textbf{S}^{ij}_{i,j+d} &=& \textbf{S}^{ij}_{j+d,i}=1,\\
  \textbf{S}^{ij}_{j,i+d} &=& \textbf{S}^{ij}_{i+d,j}=-1.
\end{eqnarray}
\end{subequations}
Second, he defined
\begin{equation}\label{cij}
 \textbf{C}_{ij}(\rho)=max(0, \lambda^{ij}_1-\lambda^{ij}_2-\lambda^{ij}_3-\lambda^{ij}_4),
\end{equation}
where the $\lambda^{ij}$, ordered decreasingly, are the square
roots of the four largest eigenvalues of the matrix $\rho
\textbf{S}^{ij} \rho^*\textbf{S}^{ij}$, $\rho$ is the density
matrix of the qubit-qudit system and $\rho^*$ is its conjugate.
Thirdly, he denoted the lower bound of the concurrence
$\textbf{C}(\rho)$ of the qubit-qudit system by
$\textbf{C}_{db}(\rho)$,
\begin{equation}\label{cdb}
\textbf{C}_{db}=\left[\sum_{j>i}\sum_{i=0}^{d-2}
\textbf{C}_{ij}^2(\rho)\right]^{1/2}\leq \textbf{C}(\rho).
\end{equation}
The desired lower bound on the qubit-qudit EOF is
$\varepsilon[C_{db}(\rho)]$.

For a qubit-qutrit mixed state system, there are only three
$\textbf{S}^{ij}$ expressed as $S_x,S_y$ and $S_z$,
\begin{equation}\label{sx}
    \textbf{S}_x=\left(%
\begin{array}{cccccc}
  0 & 0 & 0 & 0 & 1 & 0 \\
  0 & 0 & 0 &-1 & 0 & 0 \\
  0 & 0 & 0 & 0 & 0 & 0 \\
  0 &-1 & 0 & 0 & 0 & 0 \\
  1 & 0 & 0 & 0 & 0 & 0 \\
  0 & 0 & 0 & 0 & 0 & 0 \\
\end{array}%
\right),
\end{equation}

\begin{equation}\label{sy}
    \textbf{S}_y=\left(%
\begin{array}{cccccc}
  0 & 0 & 0 & 0 & 0 & 1 \\
  0 & 0 & 0 & 0 & 0 & 0 \\
  0 & 0 & 0 &-1 & 0 & 0 \\
  0 & 0 &-1 & 0 & 0 & 0 \\
  0 & 0 & 0 & 0 & 0 & 0 \\
  1 & 0 & 0 & 0 & 0 & 0 \\
\end{array}%
\right),
\end{equation}

\begin{equation}\label{sz}
    \textbf{S}_z=\left(%
\begin{array}{cccccc}
  0 & 0 & 0 & 0 & 0 & 0 \\
  0 & 0 & 0 & 0 & 0 & 1 \\
  0 & 0 & 0 & 0 &-1 & 0 \\
  0 & 0 & 0 & 0 & 0 & 0 \\
  0 & 0 &-1 & 0 & 0 & 0 \\
  0 & 1 & 0 & 0 & 0 & 0 \\
\end{array}%
\right),
\end{equation}
respectively. For the qubit-qubit mixed state, there is only one
$S^{ij}$ denoted as
\begin{equation}\label{s}
    \textbf{S}=\left(%
\begin{array}{cccc}
  0 & 0 & 0 & -1 \\
  0 & 0 & 1 & 0 \\
  0 & 1 & 0 & 0 \\
  -1 & 0 & 0 & 0 \\
\end{array}%
\right).
\end{equation}
Note that the Eq. (\ref{s}) is a matrix constructed via $\sigma^A_y\otimes\sigma^B_y$, with $A$ and $B$ denoting the indexes of qubit $A$ and qubit $B$, respectively. Considering  a mixed qubit-qubit quantum system with
levels $|0\rangle$ and $|1\rangle$ one can construct the Pauli operator
$\sigma_y$ for the subsystems $A$ and $B$ given by
\begin{equation}\label{sigma_x_y}
    \sigma^x_y=i \left(\left|1_x\rangle \langle
   0_x\right|-\left|0_x\rangle \langle
   1_x\right|\right),~~~ (x=A,B).
\end{equation}
Then, if we consider a
mixed qubit-qutrit system composed of subsystem $A$ with two
levels $|0_A\rangle$ and $|1_A\rangle$ and a subsystem $B$ with
three levels $|0_B\rangle,|1_B\rangle,|2_B\rangle$, the Pauli
operator $\sigma_y$ can be now expressed by Dirac notation as:
\begin{equation}\label{sigma_x_y_ij}
    \sigma^x_{y,ij}=i \left(\left|j_x\rangle \langle
   i_x\right|-\left|i_x\rangle \langle
   j_x\right|\right),~~~ (x=A,B),
\end{equation}
where $\{i,j\}=\{0,1\}$ for $x=A$ and $\{i,j\}=\{0,1\},\{0,2\}$
and $\{1,2\}$, respectively, for $x=B$. It is easy to test the
matrix forms of $\sigma^A_{y,01}\otimes\sigma^B_{y,01}$,
$\sigma^A_{y,01}\otimes\sigma^B_{y,02}$ and
$\sigma^A_{y,01}\otimes\sigma^B_{y,12}$ in basics
$|0_A,0_B\rangle,|0_A,1_B\rangle,
|0_A,2_B\rangle,|1_A,0_B\rangle,|1_A,1_B\rangle,|1_A,2_B\rangle$
are $-\textbf{S}_x$,~$-\textbf{S}_y$ and $-\textbf{S}_z$,
respectively.
 Generally speaking, for a qubit-qudit mixed state system
the matrix forms of $\sigma^A_{y,01}\otimes\sigma^B_{y,ij}$
$(i<j)$ in basics $|0_A,0_B\rangle,~|0_A,1_B\rangle,~
...,~|0_A,(d-1)_B\rangle,~ |1_A,0_B\rangle,~
|1_A,1_B\rangle,~...,~|1_A,(d-1)_B\rangle$ equals
$-\textbf{S}^{ij}$.
 It is notable that the difference between the
matrix expression of $\sigma^A_{y,01}\otimes\sigma^B_{y,ij}$ and
$\textbf{S}^{ij}$ is only a minus. It doesn't affect the
eigenvalues of $\rho \textbf{S}^{ij} \rho^*\textbf{S}^{ij}$ if we
replace $\textbf{S}^{ij}$ by
$\sigma^A_{y,01}\otimes\sigma^B_{y,ij}$.

It is now clear that Gerjuoy's approach, in fact, is to treat the
qubit-qudit mixed system as a set of $d!/(2! (d-2)!)$ qubit-qubit
system. Therefore, for such a system, we can take any two levels
of qudit subsystem to combine with the qubit subsystem as a
qubit-qubit. Then, all methods to solving the qubit-qubit problem
can be used. In our present case, we only need to use matrix
$\textbf{S}$ instead of  $\textbf{S}^{ij}$ (or
 $\textbf{S}_x,\textbf{S}_y$ and $\textbf{S}_z$) to calculate the
concurrence $\textbf{C}^{ij}(\textbf{C}_x,\textbf{C}_y$ and
$\textbf{C}_z)$ for those qubit-qubit subsystem. This will greatly
simplify the calculation. In the next section, we shall
demonstrate our proposal by a concrete example.

\section{ILLUSTRATIVE EXAMPLES}\label{s3}
To illustrate our method let us consider two atoms (A and B), each
of them interacting resonantly with a single quantized mode of a
cavity field (system C) in a Fock state. This physical situation
is described by the two-atom Tavis-Cummings (TC) Hamiltonian:
 $H
=\hbar g[(\sigma_A + \sigma_B)a_C^\dag + (\sigma_A^\dag +
\sigma_B^\dag)a_C]$, where $\sigma_j$ and $\sigma_j^\dag$ are the
Pauli ladder operators for the $j$th atom, $a (a^\dag)$ is the
annihilation (creation) operator for photons in cavity $C$, and
$g$ is the coupling constant. We assume the system is initially in
the state $|\psi(0)\rangle = (\alpha|0_{_A} 0_ {_B}\rangle +
\beta|1_{_A} 1_{_B}\rangle)|n_{_C}\rangle$. Since the TC
Hamiltonian preserves the total number of excitations, the cavity
mode will evolve within a five-dimensional Hilbert space spanned
by $\{|(n-2)_{_C}\rangle,|(n-1)_{_C}\rangle,|n_{_C}\rangle,|(n +
1)_{_C}\rangle,|(n + 2)_{_C}\rangle\}$ for $n\geq 2$. When $n =
0,1$ the dimension will be $3$ and $4$, respectively. On the other
hand, the atomic system will evolve within the subspace
$\{|0_{_A}0_{_B}\rangle,|+\rangle,|1_{_A}1_{_B}\rangle\}$ with
$|+\rangle = (|1_{_A}0_{_B}\rangle +
|0_{_A}1_{_B}\rangle)/\sqrt{2}$ independently of $n$. By solving
the Schr\"{o}dinger equation, the system at time $t$ is described
by the state
\begin{eqnarray}
  |\psi(t)\rangle &=& c_1(t)|0_{_A} 0_{_B}\rangle|(n + 2)_{_C}\rangle + c_2(t )|+\rangle|(n +
1)_{_C}\rangle   \nonumber \\
   && +c_3(t )|1_{_A}1_{_B}|n_{_C}\rangle +
c_4(t)|0_{_A}0_{_B}\rangle|n_{_C}\rangle\nonumber \\
   && +c_5(t)|+\rangle|(n-1)_{_C}\rangle + c_6(t
)|1_{_A}1_{_B}|(n-2)_{_C}\rangle,\nonumber \\
\end{eqnarray}
where the probability amplitudes are

\begin{equation}\label{c1}
   c_1(t)=-\frac{\beta \sqrt{(n+1)(n+2)}}{2 n+3}[1-\cos(\sqrt{2 (2n+3)}g
     t)],
\end{equation}
\begin{equation}\label{c2}
    c_2(t)= -\frac{i \beta \sqrt{n+1}}{\sqrt{2 n+3}}\sin(\sqrt{2 (2n+3)}g
     t),
\end{equation}
\begin{equation}\label{c3}
    c_3(t)=\beta \left\{1-\frac{n+1}{2 n+3}[1-\cos(\sqrt{2 (2n+3)}g
     t)]\right\},
\end{equation}
\begin{equation}\label{c4}
    c_4(t) =  \alpha \left\{1-\frac{n}{2 n-1}[1-\cos(\sqrt{2 (2n-1)}g
     t)]\right\},
\end{equation}
\begin{equation}\label{c5}
    c_5(t) = -\frac{i \alpha \sqrt{n}}{\sqrt{2n-1}}\sin(\sqrt{2 (2n-1)}g
     t),
\end{equation}
\begin{equation}\label{c6}
    c_6(t) = -\frac{\alpha \sqrt{n(n-1)}}{2 n-1}[1-\cos(\sqrt{2(2n-1)}g t)].
\end{equation}

Now, we take trace of density operator
$\rho=|\psi(t)\rangle\langle \psi(t)|$ over atom $B$ resulting in
the reduced density operator of the qubit-qudit system
$\rho_{AC}$.
\subsection{Qubit-Qutrit case}
When $n=0$, atom A and cavity $C$ compose a qubit-qutrit system.
As described in the above section, we delete terms not containing
$|i_A,0_C\rangle \langle j_A,0_C|$,$|i_A,0_C\rangle \langle
j_A,1_C|$, $|i_A,1_C\rangle \langle j_A,0_c|$ and $|i_A,1_C\rangle
\langle j_A,1_C|~(i,j=0,1)$ in $\rho_{AC}$ to form
\begin{eqnarray}\label{rhoAC01}
  \rho_{AC}^{01} &=& c_3 c_3^* |1_A,0_C\rangle \langle 1_A,0_C|
  +c_4 c_4^* |0_A,0_C\rangle \langle 0_A,0_C|\nonumber\\
   &&+\frac{1}{2}c_2c_2^*(|0_A,1_C\rangle \langle 0_A,1_C|
  + |1_A,1_C\rangle \langle 1_A,1_C|)\nonumber  \\
   &&+ \frac{1}{\sqrt{2}}(c_2 c_4^* |1_A,1_C\rangle \langle 0_A,0_C|
  +c_2^* c_4 |0_A,0_C\rangle \langle 1_A,1_C|)\nonumber \\
   && +\frac{1}{\sqrt{2}}(c_3 c_2^* |1_A,0_C\rangle \langle 0_A,1_C|
  +c_2^* c_3 |0_A,1_C\rangle \langle 1_A,0_C|),\nonumber\\
\end{eqnarray}
The matrix form of which is
\begin{equation}\label{rhoAC01m}
  \rho_{AC}^{01} = \left(
\begin{array}{cccc}
  c_4 c_4^* & 0 & 0 &c_4 c_2^*/\sqrt{2} \\
 0          & c_2 c_2^*/2 & c_2 c_3^*/\sqrt{2} & 0 \\
  0 & c_3 c_2^*/\sqrt{2} & c_3 c_3^* & 0 \\
  c_2 c_4^*/\sqrt{2} & 0 & 0 & c_2 c_2^*/2 \\
\end{array}
\right).
\end{equation}
Similarly we obtain matrices $\rho_{AC}^{02} $ and $\rho_{AC}^{12}
$ respectively
\begin{equation}\label{rhoAC02m}
  \rho_{AC}^{02} = \left(
\begin{array}{cccc}
c_4 c_4^* & c_4 c_1^* & 0 & 0\\
c_1 c_4^* & c_1 c_1^* & 0 & 0\\
0 & 0 & c_3 c_3^* & 0\\
0 & 0 & 0 & 0\\
\end{array}
\right),
\end{equation}
\begin{equation}\label{rhoAC12m}
  \rho_{AC}^{12} = \left(
\begin{array}{cccc}
c_2 c_2^*/2 & 0 & 0 & 0\\
0 &  c_1 c_1^* & c_1 c_2^*/\sqrt{2} & 0\\
0 & c_2 c_1^*/\sqrt{2} & c_2 c_2^*/2 & 0\\
0 & 0 & 0 & 0\\
\end{array}
\right).
\end{equation}
The matrices $\rho_{AC}^{01}$ and $\rho_{AC}^{12}$ are $X$
form~\cite{xform}. The corresponding concurrences can be read out
\begin{equation}\label{cx}
    \textbf{C}_x= \textbf{C}_{01}=\sqrt{2} \left|\left|c_2 c_4\right|-\left|c_2
   c_3\right|\right|,
\end{equation}
\begin{equation}\label{cz}
    \textbf{C}_z= \textbf{C}_{12}=\sqrt{2} \left|c_1 c_2\right|.
\end{equation}
Unfortunately, The matrix $\rho_{AC}^{02}$ is not the $X$ form.
The square roots of four eigenvalues of $\rho_{AC}^{02}
\textbf{S}\rho_{AC}^{02*}\textbf{S}$ are
$\{0, 0, |c1 c3|, |c1 c3|\}$, therefore,
$\textbf{C}_y=\textbf{C}_{02}=0$. Consequently, the lower bound of
concurrence $\textbf{C}(\rho)$ of this qubit-qutrit system is
$C_{AC}=\sqrt{2[\left|c_1 c_2\right|^2
              +(\left|c_2 c_4\right|-\left|c_2 c_3\right|)^2]}$.
The lower bound of EOF for the qubit-qutrit system is
$E_f(C_{AC})=h((1+\sqrt{1-C_{AC}^2})/2)$, where $h(x)=-x
\log_2x-(1-x)\log(1-x)$ is double entropy function. The evolution
of $E_{AC}$ with the dimensionless time $\tau=\sqrt{6}gt/(2 \pi)$
are plotted in Fig.1. In order to make a comparison, $E_{AC}$ also
was computed according to the lower bound of $C_{AC}$ given by
Ref.~\cite{C7}.
Though two lines in Fig.1 do not coincide, but their behaviors of
evolution with time are the same.

\begin{figure}[tb]
\centering
\includegraphics[width=6.5cm]{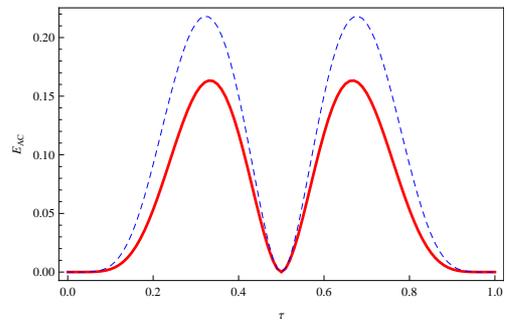}
\caption{(Color online) The evolution of entanglement of formation
between the atom $A$ and the cavity mode $C$ for the initial state
$|\psi(0)\rangle = (\alpha|0_{_A} 0_ {_B}\rangle + \beta|1_{_A}
1_{_B}\rangle)|0_{_C}\rangle$ with $\alpha =\beta = 1/\sqrt{2}$.
The red solid line corresponds to the present expression of
$\textbf{C}_{AC}$ and the blue dashed line to the lower bound of
$\textbf{C}_{AC}$ given by Ref.~\cite{C7}. The dimensionless time
$\tau=\sqrt{6}gt/(2 \pi)$.} \label{F1}
\end{figure}

\subsection{Qubit-Qudit case}

Now, considering $n=2$, the atom $A$ and cavity $C$ compose a qubit-qudit $(d=5)$
system. According to our scheme, we can take any two levels of
cavity mode $C$ to form ten qubit-qubit system with two levels of
atom $A$ and find:
\begin{equation}\label{r01}
   \rho_{AC}^{01}=\left(%
\begin{array}{cccc}
 0 & 0 & 0 & 0\\
 0 & \frac{c_5 c_5^*}{2} & \frac{c_5 c_6^*}{\sqrt{2}} & 0\\
 0 & \frac{c_6 c_5^*}{\sqrt{2}} & c_6 c_6^* & 0\\
 0 & 0 & 0 & \frac{c_5 c_5^*}{2}\\
\end{array}%
\right),
\end{equation}
\begin{equation}\label{r02}
     \rho_{AC}^{02}=\left(%
\begin{array}{cccc}
 0 & 0 & 0 & 0\\
 0 & c_4 c_4^* & 0 & 0\\
 0 & 0 & c_6 c_6^* & c_6 c_3^*\\
 0 & 0  & c_3 c_6^* & c_3 c_3^*\\
\end{array}%
\right),
\end{equation}
\begin{equation}\label{r03}
     \rho_{AC}^{03}=\left(%
\begin{array}{cccc}
0 & 0 & 0 & 0\\
0 & \frac{c_2 c_2^*}{2}, \frac{c_2 c_6^*}{\sqrt{2}} &  0\\
0  & \frac{c_6 c_2^*}{\sqrt{2}} & c_6 c_6^* & 0\\
0 & 0 & 0 & \frac{c_2 c_2^*}{2}\\
\end{array}%
\right),
\end{equation}
\begin{equation}\label{r04}
     \rho_{AC}^{04}=\left(%
\begin{array}{cccc}
0 & 0 & 0 & 0\\
0 & c_1 c_1^* & 0 & 0\\
0 &  0 & c_6 c_6^* & 0\\
0 & 0 & 0 & 0\\
\end{array}%
\right),
\end{equation}
\begin{equation}\label{r12}
   \rho_{AC}^{12}=\left(%
\begin{array}{cccc}
\frac{c_5 c_5^*}{2} & 0 & 0 & \frac{c_5 c_3^*}{\sqrt{2}}\\
0 & c_4 c_4^* & \frac{c_4 c_5^*}{\sqrt{2}} & 0\\
0 & \frac{c_5 c_4^*}{\sqrt{2}} & \frac{c_5 c_5^*}{2} & 0\\
\frac{c_3 c_5^*}{\sqrt{2}} &  0 & 0 & c_3 c_3^*\\
\end{array}%
\right),
\end{equation}
\begin{equation}\label{r13}
    \rho_{AC}^{13}=\left(%
\begin{array}{cccc}
\frac{c_5 c_5^*}{2} & \frac{c_5 c_2^*}{2} &  0 & 0\\
  \frac{c_2 c_5^*}{2} & \frac{c_2 c_2^*}{2} & 0 & 0\\
  0  & 0 &  \frac{c_5 c_5^*}{2} &  \frac{c_5 c_2^*}{2}\\
  0 & 0 &  \frac{c_2 c_5^*}{2} & \frac{c_2 c_2^*}{2}\\
  \end{array}%
\right),
\end{equation}

\begin{equation}\label{r14}
    \rho_{AC}^{14}=\left(%
\begin{array}{cccc}
\frac{c_5c_5^*}{2} & 0 & 0 & 0\\
0  & c_1 c_1^* &, \frac{c_1 c_5^*}{\sqrt{2}} &  0\\
0  & \frac{c_5 c_1^*}{\sqrt{2}} & \frac{c_5 c_5^*}{2} & 0\\
0 & 0 & 0 & 0\\
  \end{array}%
\right)
\end{equation}
\newpage
\begin{equation}\label{r23}
   \rho_{AC}^{23}=\left(
\begin{array}{cccc}
 c_4 c_4^* & 0 & 0 & \frac{c_4 c_2^*}{\sqrt{2}} \\
 0 & \frac{c_2 c_2^*}{2} & \frac{c_2 c_3^*}{\sqrt{2}} & 0 \\
 0 & \frac{c_3 c_2^*}{\sqrt{2}} & c_3 c_3^* & 0 \\
 \frac{c_2 c_4^*}{\sqrt{2}} & 0 & 0 & \frac{c_2 c_2^*}{2}
 \end{array}%
 \right),
\end{equation}
\begin{equation}\label{r24}
     \rho_{AC}^{24}=\left(
\begin{array}{cccc}
c_4 c_4^* & c_4c_1^* & 0 & 0 \\
c_1 c_4^* & c_1c_1^* & 0 & 0 \\
 0 & 0 & c_3 c_3^* & 0 \\
 0 & 0 & 0 & 0\\
 \end{array}%
 \right),
\end{equation}
\begin{equation}\label{r34}
   \rho_{AC}^{34}=\left(
\begin{array}{cccc}
 \frac{\text{c2} \text{c2}^*}{2} & 0 & 0 & 0 \\
 0 & \text{c1} \text{c1}^* & \frac{\text{c1}
   \text{c2}^*}{\sqrt{2}} & 0 \\
 0 & \frac{\text{c2} \text{c1}^*}{\sqrt{2}} &
   \frac{\text{c2} \text{c2}^*}{2} & 0 \\
 0 & 0 & 0 & 0
 \end{array}%
 \right).
\end{equation}
The eigenvalues of $\rho_{AC}^{ij}
\textbf{S}\rho_{AC}^{ij*}\textbf{S}$ are all equal to that of
$\rho_{AC}\textbf{S}^{ij}\rho_{AC}^*\textbf{S}^{ij}~(i=0, ... 3,
~j>i)$, The non-zero square roots of them are $ \{\sqrt{2} |c_5
c_6|\},
 \{|c_4 c_6|,|c_4 c_6|\},
 \{\sqrt{2} |c_2 c_6|\},
  \{|c_1 c_6|,c_1 c_6|\},$
   $\{\sqrt{2} |c_3 c_5|,\sqrt{2}
   |c_4 c_5|\},
   \{\sqrt{2} |c_1 c_5|\},
\{\sqrt{2} |c_2 c_3,\sqrt{2}|c_2 c_4|\},$ $ \{|c_1 c_3|,|c_1
c_3|\},\{\sqrt{2}|c_1 c_2|\}$, respectively. Correspondingly,
$C_{01} = \sqrt{2} |c_5 c_6|, C_{03}= \sqrt{2} |c_2
c_6|,C_{12}=\sqrt{2} ||c_3 c_5|-|c_4
c_5||,C_{14}=\sqrt{2}|c_1c_5|,C_{23}=\sqrt{2} ||c_2 c_3|-|c_2
c_4||,C_{34}=\sqrt{2} |c_1 c_2|,C_{02}=C_{04}=C_{13}=C_{24}=0.$
The lower bound of concurrence $\textbf{C}(\rho)$ of the
qubit-qudit is
\begin{widetext}
\begin{equation}\label{cbd}
C_{b5}=\sqrt{2} \sqrt{(|c_2 c_3|-|c_2 c_4|)^2+(|c_3
   c_5|-|c_4 c_5|)^2+c_1^2
   c_2^2+c_1^2 c_5^2+c_2^2 c_6^2+c_5^2 c_6^2}.
\end{equation}
\end{widetext}
We plot the evolution of $E_{AC}$ with the dimensionless time
$\tau=\sqrt{14}gt/(6 \pi)$ in Fig. 2. $E_{AC}$ is also computed
according to the lower bound of $C_{AC}$ given by Ref.~\cite{C7}
 and plotted in the same figure for comparison. Though two lines in
Fig. 2 have different trends in some interval of $\tau$, their
global behaviors of evolution with time are basically the same.

\begin{figure}[tb]
\centering
\includegraphics[width=6.5cm]{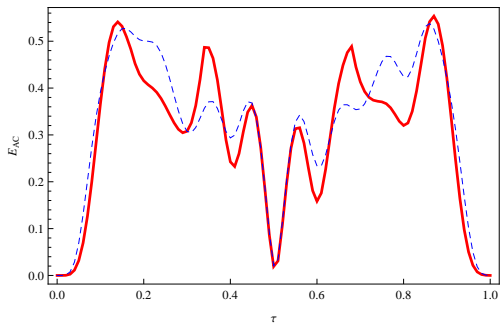}
\caption{(Color online) The evolution of entanglement of formation
between the atom $A$ and the cavity mode $C$ for the initial state
$|\psi(0)\rangle = (\alpha|0_{_A} 0_ {_B}\rangle + \beta|1_{_A}
1_{_B}\rangle)|2_{_C}\rangle$ with $\alpha =\beta = 1/\sqrt{2}$.
The red solid line corresponds to the present expression of
$\textbf{C}_{AC}$ and the blue dashed line to the lower bound of
$\textbf{C}_{AC}$ given by Ref.~\cite{C7}. The dimensionless time
$\tau=\sqrt{14}gt/(6 \pi)$.} \label{F2}
\end{figure}

\section{SUMMARY}\label{s4}
We have analyzed Gerjuoy's approach on calculating the lower bound
on entanglement of formation  for qubit-qudit system and we found
that his method, in fact, is to treat qubit-qudit system as a set
of qubit-qubit system. Therefore, we proposed a simple scheme to
solve qubit-qudit problem. The scheme consists of three steps: (1)
partition the qudit system into a set of qubit system; (2) compose
the original qubit and partitioned qubit into a set of qubit-qubit
systems and treat them by all methods suitable to qubit-qubit
system. Find the measurements you want for every qubit-qubit
system; (3) obtain the measurement of whole qubit-qudit system.
For the case discussed in the present paper, we calculated the
concurrences for every qubit-qubit system and the lower bound of
the concurrence of the qubit-qutrit or qubit-qudit system. Our
method has the advantage of avoiding finding many matrices
$\textbf{S}^{ij}$ and only using one matrix
$\textbf{S}=\sigma_y\otimes\sigma_y$. This method greatly
simplified the calculation about the measurement of qubit-qudit
system. We hope this method can be extended to treat other
problems of the qubit-qudit system.

\begin{acknowledgments}
This work is supported by the Special Funds  for Theoretical
Physics  of the National Natural Science Foundation of China
(Grant No.11147161). The partial support by the CNPq and INCT-IQ (WBC) are also
acknowledged.
\end{acknowledgments}

\end{document}